\documentclass[aps,prl,superscriptaddress, twocolumn, showpacs]{revtex4-1}
\usepackage{graphicx}
\usepackage{dcolumn}
\usepackage{bm}
\usepackage{amsmath}

\usepackage{color}
\usepackage[flushleft]{threeparttable}
\usepackage{bm,graphicx,hyperref}
\hypersetup{%
  breaklinks = {true},
  citecolor = {blue},
  colorlinks = {true},
  linkcolor = {red},
}

\begin{document}

\title{Polarization and Valley Switching in Monolayer Group-IV
  Monochalcogenides}

\author{Paul~Z.~Hanakata}
\affiliation{Department of Physics, Boston University, Boston, MA 
02215}

\author{Alexandra~Carvalho} \affiliation{Centre for Advanced 2D
  Materials and Graphene Research Centre, National University of
  Singapore, 6 Science Drive 2, 117546, Singapore}

\author{David~K.~Campbell\footnote{Electronic address: dkcampbe@bu.edu}}
\email[Corresponding author: ]{dkcampbe@bu.edu}
\affiliation{Department of Physics, Boston University, Boston, MA  02215}

\author{Harold~S.~Park\footnote{Electronic address: parkhs@bu.edu}}
\email[Corresponding author: ]{parkhs@bu.edu}
\affiliation{Department of Mechanical Engineering, Boston University, Boston, MA 
02215}

\date{\today}
\begin{abstract}
  Group-IV monochalcogenides are a family of two-dimensional puckered
  materials with an orthorhombic structure that is comprised of polar
  layers.  In this article, we use first principles calculations to
  show the multistability of monolayer SnS and GeSe, two prototype
  materials where the direction of the puckering can be switched by
  application of tensile stress or electric field.  Furthermore, the
  two inequivalent valleys in momentum space, which dictated by the
  puckering orientation, can be excited selectively using linearly
  polarized light, and this provides additional tool to identify the
  polarization direction.  Our findings suggest that SnS and GeSe
  monolayers may have observable ferroelectricity and multistability,
  with potential applications in information storage.
\end{abstract}


\maketitle
The discovery of 2D materials that can be isolated into single layers
through exfoliation and exhibit novel properties has established new
paradigms for ultrathin devices based on atomically sharp
interfaces~\cite{exfo-CSR2014, exfo-SCIE2013}. In particular,
transition metal dichalcogenides (TMDs) have been studied extensively
and have shown potential for many technological applications ranging
from photovoltaics to valleytronic devices~\cite{wangNNANO2012,
  johariACSNANO2012, chhowallaNC2013, zengNN2012,
  xiao-PRL-108-196802-2012, sieNM14, kimSCIE14}.  The family of
monolayer 2D materials has recently grown to include other 2D
semiconductors, such as phosphorene and related materials.

However, one of the features thus far lacking for applications both in
2D electronics and in valleytronics is non-volatile memory.
Ferromagnetism, an essential element in spintronic memories, is
believed to be achievable in graphene and other 2D materials but so
far remains difficult to realize and
control~\cite{sepioni-PRL-105-207205-2010}.  Ferroelectric memories,
in which the information is stored in the orientation of the electric
dipole rather than in the magnetization are a possible option.
Single-layer graphene (SLG) ferroelectric field-effect transistors
(FFET) with symmetrical bit writing have been
demonstrated~\cite{song-APL-99-042109-2011}, but the prototypes rely
on bulk or thin film ferroelectric
substrates~\cite{song-APL-99-042109-2011} or ferroelectric
polymers~\cite{zheng-PRL-105-166602-2010}, rather than on crystalline
atomically thin ferroelectric materials.  An altogether different
approach to information storage relies on phase change materials,
where the bit value corresponds to a distinct structural phase of the
material.  Researchers have recently optimized the phase switching
energy by using superlattice structures where the movement of the
atoms is confined to only one dimension~\cite{simpson-NN-6-501-2011}.

In this article, we analyze the stability of group-IV monochalcogenide
MX (M=Ge or Sn, and X=S or Se) monolayers, paying particular interest
to their potential as memory functional materials.  As prototypes, we
use SnS and GeSe. In ambient conditions, bulk SnS and GeSe crystallize
in the orthorhombic structure of the $Pnma$ space group.  At 878 K,
SnS goes through a second-order displacive phase transition into the
$\beta$-SnS phase with $Cmcm$ symmetry~\cite{alptekinJMM2011,
  chattopadhyay1984temperature}, which is also a layered phase that
can be viewed as a distorted rocksalt structure.  For bulk GeSe, such
a phase transition has not been observed.  Instead, at 924 K bulk GeSe
transforms into the rocksalt phase ($Fm\bar{3}m$).  This phase can
also be stabilized using external pressure~\cite{deringerPRB2014}.

Similar to phosphorene~\cite{rodinPRL2014,liuACSN2014}, $Pnma$ SnS and
GeSe can be exfoliated~\cite{jack-ACS-137-12689-2015,
  mukherjee-ACS-5-9594-2013}. In monolayer form, they feature multiple
valleys, large spin-orbit splitting\cite{gomesPRB2015} and a
piezoelectric coefficient that surpasses that of the
TMDs~\cite{feiAPL2015, gomes-PRB-92-214103-2015}.  Having an in-plane
polar axis makes SnS and GeSe monolayers capable of a mechanical
response to an applied electric field.

Here, we use density functional theory (DFT) calculations to
characterize the multistability of SnS and GeSe, exploring ways in
which the phase transitions and domain switch can be triggered
externally.  We start by demonstrating how the reversible phase
transition can be induced by uniaxial stress or electric field.  Then,
we show how the phase and lattice orientation states can be detected
using the valley properties.

\section{Methods}
The calculations were based on density functional theory (DFT)
implemented in the {\sc Quantum ESPRESSO} package~\cite{QE-2009}.  The
generalized gradient approximation (GGA) of Perdew-Burke-Ernzerhof
(PBE) was used for the exchange and correlation functional, and
Troullier-Martins type pseudopotentials~\cite{troullierPRB1991}.  The
Kohn-Sham orbitals were expanded in a plane-wave basis with a cutoff
energy of 70 Ry, and for the charge density a cutoff of 280 Ry was
used.  A $k$-point grid sampling grid was generated using the
Monkhorst-Pack scheme with 10$\times$10$\times$1
points~\cite{monkhorstPRB1976}, and a finer regular grid of
80$\times$80$\times$1 was used for transition probability
calculations.  The equilibrium structures were found by using a
conjugate-gradient optimization algorithm, and the energy landscape is
mapped by relaxing the structure under constraints for each of the
in-plane lattice parameters, while all the other structural parameters
are allowed to relax.

We used the modern theory of
polarization~\cite{vanderbilt-PRB-47-1651-1993} to calculate the
spontaneous polarization given by
\begin{equation} 
\vec{\mathcal{P}}=\frac{1}{\Omega}\sum_{\tau}q^{\rm{ion}}_{\tau}{\bf R}_{\tau}-\frac{2i\rm{e}}{(2\pi)^3}\sum_{n}^{\rm{occ}}\int_{BZ}d^3{\bf k}e^{-i\vec{k}\cdot{\bf R}}\Big\langle u_{n{\bf k}}\Big|\frac{\partial u_{n{\bf k}}}{\partial {\bf k}}\Big\rangle, 
\label{eq:pol}
\end{equation}
where $q_\tau$ is the ionic charge plus the core electrons,
${\bf R}_{\tau}$ is the position of ions, $\Omega$ is the unit cell
volume, $\rm{e}$ is the elementary charge, $n$ is the valence band
index, ${\bf k}$ is the wave vector, and $u_{n{\bf k}}$ is the
electronic wave function. The first term is the contribution from ions
and core electrons, and the second term is the electronic contribution
defined as adiabatic flow of current which can be calculated from the
Berry connection~\cite{vanderbilt-PRB-47-1651-1993}.The response of
the material to a homogenous static external electric field is
calculated based on methods developed by
Refs.~\cite{souza-PRL-89-117602, umari-PRL-89-157602} implemented in
the {\sc Quantum ESPRESSO} package~\cite{QE-2009}.

\begin{figure}
\includegraphics[width=8cm]{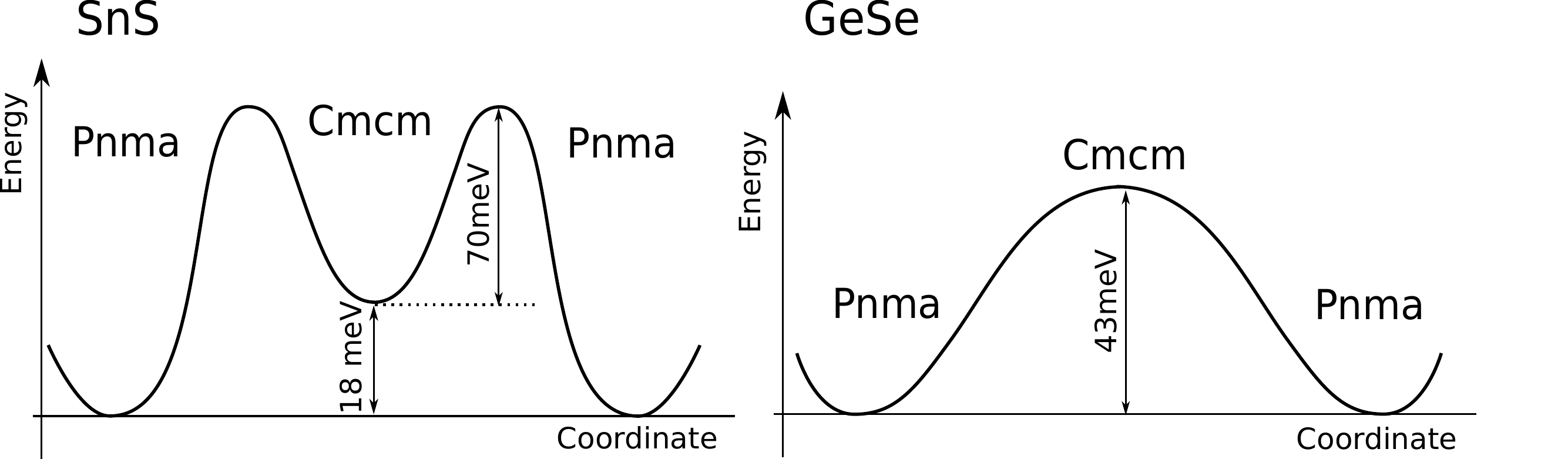}
\caption{Schematic configuration-coordinate diagram for
$Cmcm$-ML and $Pnma$-ML phases, in SnS and GeSe.} 
\label{fig:cc}
\end{figure}
\section{Results}
\subsection{Multistability of SnS and GeSe in the monolayer phase}
We start by exploring the energy landscape of monolayer SnS and
GeSe. We consider the monolayer form of the two structures that are
known for bulk SnS ie., a centrosymmetric structure ($Cmcm$), and the
$Pnma$ structure resembling black phosphorus, and which is the only
known layered structure of bulk GeSe.  We will designate the respective
monolayer phases by appending `ML' to the respective bulk space group.

\begin{figure}
  \centering
  \includegraphics[width=8.6cm]{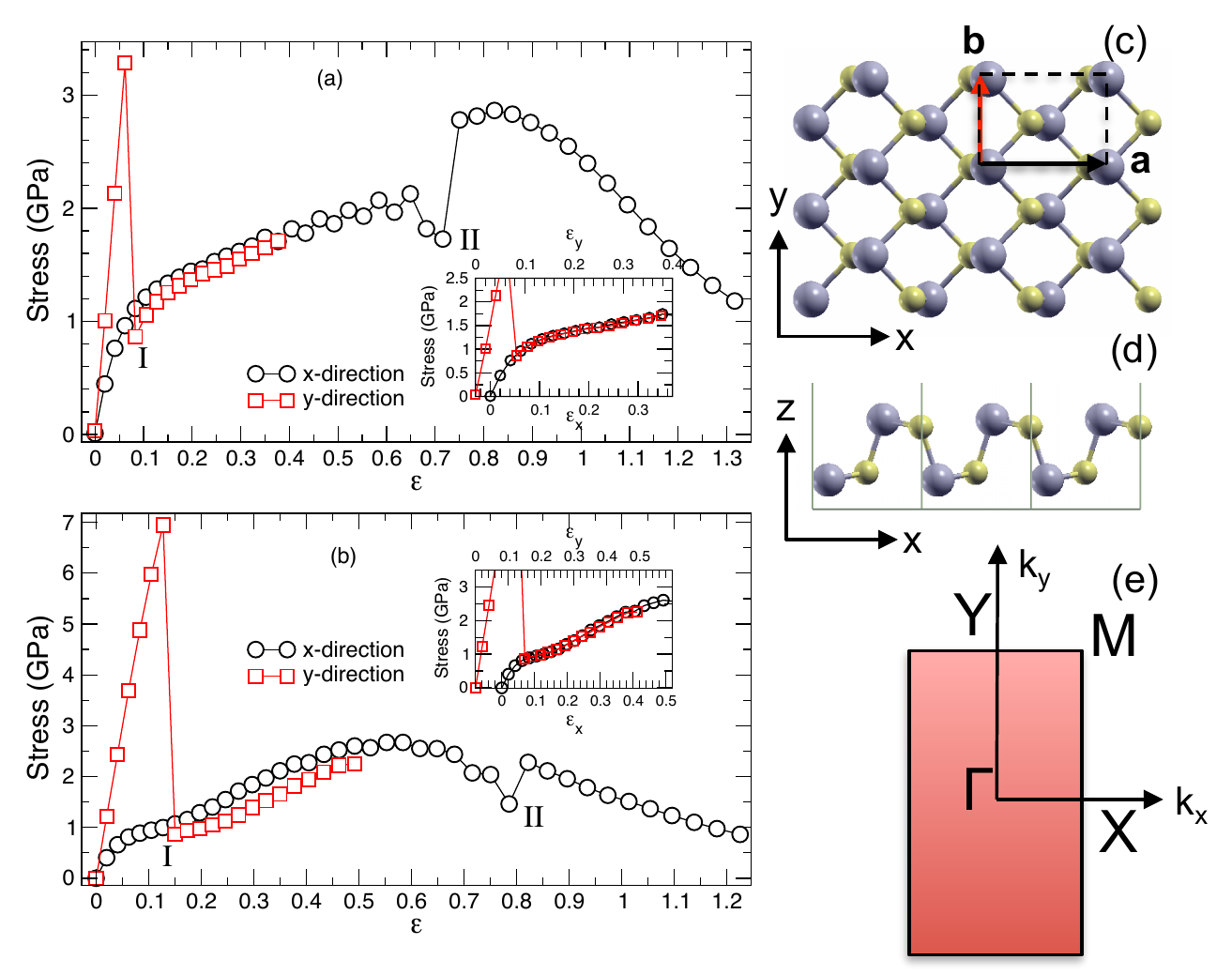}
  \caption{Stress-strain curves of monolayer (a) SnS and (b) GeSe
    for tensile strain along the $\hat{x}$ (black circle) and $\hat{y}$ (red square) directions.
    (I) indicates the $Pnma$-ML structure reconfiguration such that
    the puckering (armchair) direction $\hat{d}_{\rm puck}$ becomes
    $\hat{y}$ instead of $\hat{x}$.  (II) indicates the transformation
    into an hexagonal phase. In the insets of (a) and (b), the strain
    in the $\hat{y}$ direction was shifted to highlight the rotation
    of the $Pnma$-ML structure by $\pi/2$, swapping the armchair and
    zigzag directions. (c) and (d) top and side view of SnS structure
    with $\hat{d}_{\rm puck}=\hat{x}$. The larger grey atom is Sn and
    the smaller yellow atom is S. (e) The respective Brillouin zone
    and the high symmetry points. }
  \label{fig:stress-strain}
\end{figure}

The atomic positions in the $Pnma$-ML phase are
$\pm$(M:0.25$\pm\delta$, 0.25, 0.05; X:0.25, 0.25, -0.05) in fractional
coordinates, where M=(Sn, Ge) and X=(S, Se), $\delta=0.06$ and 0.08
for SnS and GeSe, respectively. The $Cmcm$-ML phase is obtained by
taking $\delta=0$.  As a result, the $Cmcm$-ML has two perpendicular
mirror symmetry planes, as well as inversion symmetry, while $Pnma$-ML
has no inversion symmetry. In our DFT simulations we used
$\delta=0.01$ as a tolerance to distinguish the $Pnma$-ML phase from
the $Cmcm$-ML phase ~\footnote{We define puckering orientation as a
  unit vector of the in plane bond formed by the nearest neighbor of
  MX atoms in the direction of the broken mirror symmetry. For the SnS
  structure shown in Fig.~\ref{fig:stress-strain}(d), the puckering
  direction $\hat{d}_{\rm puck}=\hat{x}$ as we define
  $\hat{d}_{\rm puck}=\frac{\vec{x_{\rm S}}-\vec{x_{\rm
        Sn}}}{|\vec{x_{\rm S}}-\vec{x_{\rm Sn}}|}$.}.

The $Pnma$-ML and the $Cmcm$-ML phases can both be seen as distortions
of a rocksalt bilayer that can be transformed into each other by a
displacement of some of the atoms along $\hat{x}$ (see
Fig.\ref{fig:stress-strain} for $\hat{x}$ and $\hat{y}$ directions).
The $Cmcm$-ML and $Pnma$-ML phases of SnS and GeSe monolayer have also
been reported in Ref.~\cite{mehboudi-NL-5b04613-2016}.  By symmetry,
there are four distinct $Pnma$-ML configurations (equivalent by
$\pi/2$ rotations of the puckering direction).  For SnS, $Cmcm$-ML is
a local minima of the energy surface.  For GeSe, the $Cmcm$-ML
structure is not an energy minimum but a saddle point.  The activation
energy for reorientation of the $Pnma$-ML puckering direction is very
small (88 meV for SnS and 43 meV for GeSe).  We note that GGA has been
successful in predicting the small enthalpy differences (tens of meV)
between different phases of ferroelectric materials, because
systematic errors cancel out when comparing systems with very similar
structures~\cite{sanna-PRB-83-054112-2011}. The broken inversion
symmetry and total energy with a typical double-well potential of SnS
and GeSe are the first two indications that these materials may
possess ferroelectricity.

\subsection{Application of uniaxial stress}
The phase transition of SnS to $Cmcm$-ML, or equivalently the
reorientation of the $Pnma$-ML structure, can be induced by in-plane
uniaxial tensile stress (Fig.~\ref{fig:stress-strain}).  We use an effective
thickness to estimate the values of stress, as outlined in
Ref.~\cite{gomes-PRB-92-214103-2015}.

For uniaxial stress along $\hat{y}$, the SnS structure begins to
resemble $Cmcm$-ML as the shorter lattice parameter $b$ is stretched.
For $\epsilon_{y}>0.08$, uniaxial stress results in the rotation of
the $Pnma$-ML structure by $\pi/2$. The puckering
$\hat{d}_{\rm puck}$thus rotates from $\hat{x}$ to $\hat{y}$
[Fig.~\ref{fig:stress-strain}(a), transition I]. Similar qualitative
behavior is observed in GeSe (see
Fig.~\ref{fig:stress-strain}(b)). Both SnS and GeSe transit to $Cmcm$
phase, but they spontaneously revert back to $Pnma$ once the tensile
stress is removed~\footnote{ We found that once the stretch is
  removed the $Cmcm$ structure spontaneously reverts back to $Pnma$
  for both SnS and GeSe. However, we found that during {\it
    compression}, the $Cmcm$ phase of SnS is stable.}.

\begin{figure*}
  \centering
  \includegraphics[width=17.2cm]{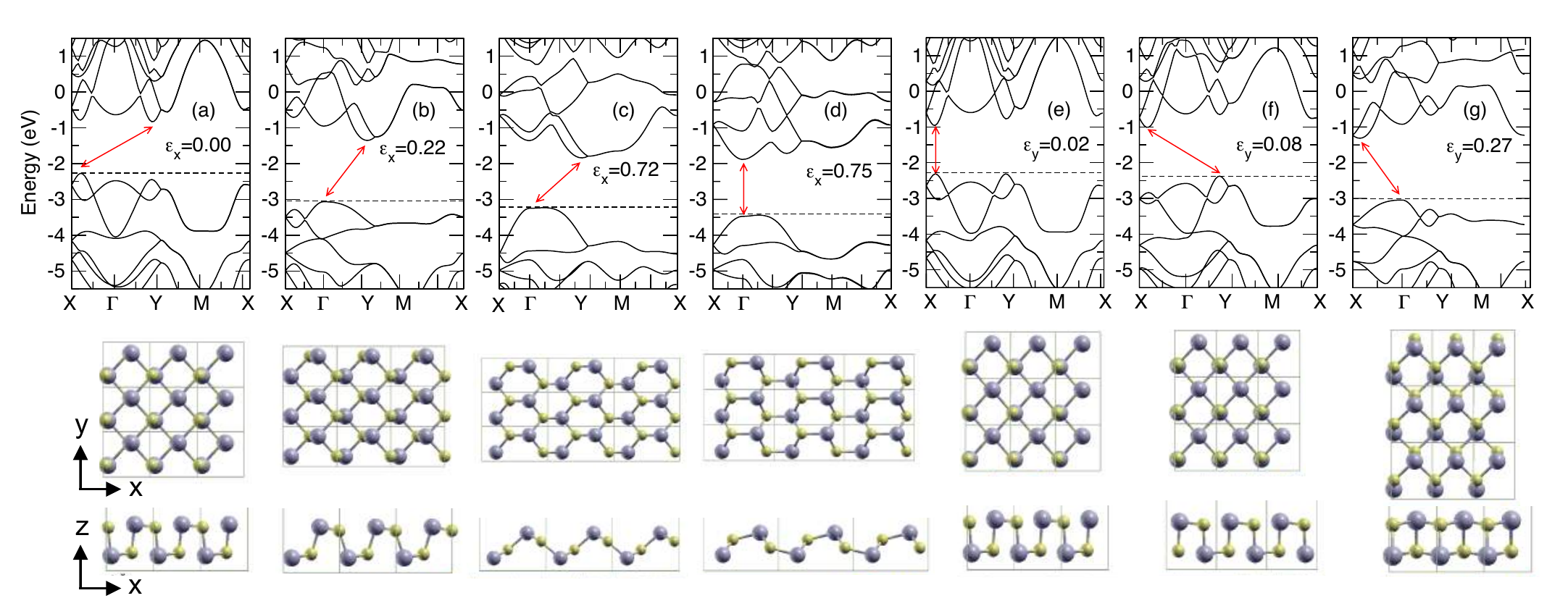}
  \caption{
    Representative band structures of SnS monolayers (a)
    unstrained, (b) to (d) under tensile uniaxial stress along the $\hat{x}$
    for axial strains of $\epsilon_{x}=0.22$ to $\epsilon_{x}=0.75$,
    and (e) to (g) under tensile uniaxial stress along $\hat{y}$ for axial
    strains of $\epsilon_{y}=0.02$ to $\epsilon_{y}=0.27$. The dotted
    lines locate the valence band maxima. The corresponding side and
    top view of structural visualizations are below the band structure
    plots.  It is apparent that the band structure (b)
    $\epsilon_x=0.22$ (or an uniaxial stress of
    $\sigma_{xx}\sim1.4$GPa) is equivalent to the band structure (g)
    $\epsilon_y=0.27$ (or an uniaxial stress of
    $\sigma_{yy}\sim1.4$GPa) if the $\hat{x}$ and $\hat{y}$ are
    inverted (rotation around $\Gamma$ axis on figures).}
  \label{fig:band1}
\end{figure*}

The application of uniaxial stress along $\hat{x}$ reveals another
phase transition at $\epsilon_{x}=0.72$ and 0.78 for SnS and GeSe,
respectively.  The structure is a hexagonal phase resembling blue
phosphorene (see Ref.~\cite{mehboudiPNAS2015}).  The hexagonal
structure and its band structure are plotted in Fig.~\ref{fig:band1}.

\begin{figure}
  \centering
  \includegraphics[width=8cm]{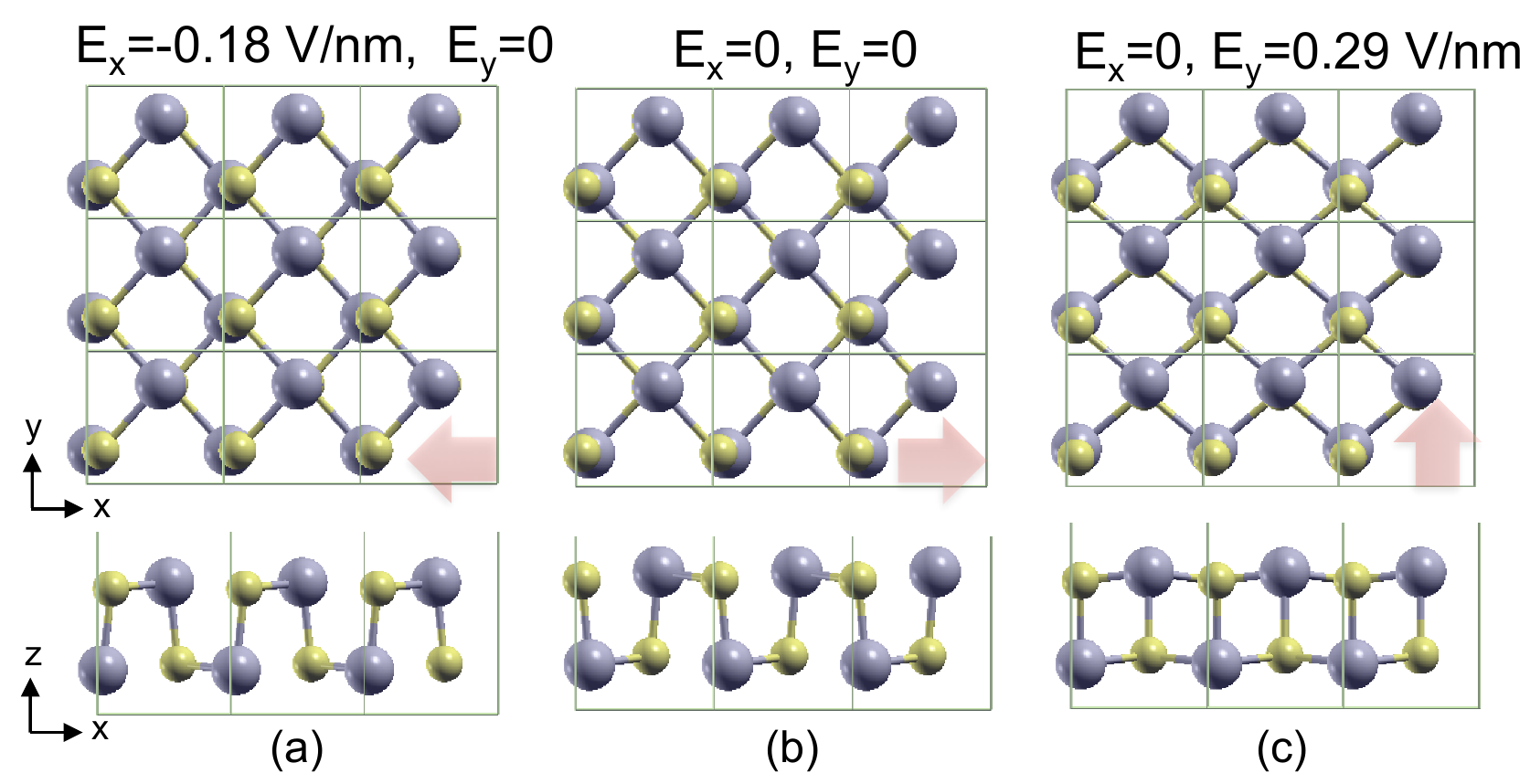}
  \caption{Structural visualization of clamped SnS monolayer under
    uniform electric field at points of transition. Puckering and
    electric dipole orientation (red arrow) can switch from positive
    $\hat{x}$ (b) to either negative $\hat{x}$ (a) or positive
    $\hat{y}$ (c) depending on the directions of applied electric
    field.  }
  \label{fig:efield}
\end{figure}

\subsection{Application of electric field}

Application of an electric field is an alternative way to trigger the
transition between different minima on the energy surface of SnS or
GeSe.  Since the $Pnma$-ML structure is piezoelectric, the application
of an electric field along the polar ($\hat{x}$) direction in a
mechanically free sample induces strain as
well~\cite{gomes-PRB-92-214103-2015}.  However, here we will consider,
for simplicity, the application of an electric field to a mechanically
clamped sample. 

The spontaneous polarization in the $Pnma$-ML phase, which was
measured with respect to the centrosymmetric structure by taking as
the effective volume the equivalent volume occupied by a layer of the
bulk unit cell, is 0.6 and 1.7 C/m$^2$ for SnS and GeSe, respectively,
which is comparable to that of 3D
ferroelectrics~\cite{zhong-PRL-72-3618-1994}.

In this case, application of an electric field with polarity opposed
to the bond dipole results in bonds breaking and creates new bonds
with inversion of the polarization along $\hat{x}$, rather than in a
rotation of the structure. As shown in Fig.~\ref{fig:efield} (a) the
ionic configuration changes (i.e., $\hat{d}_{\rm puck}$ switches from
$\hat{x}$ to $-\hat{x}$), and it is apparent from Eq.~\ref{eq:pol}
that the electric dipole orientation can be switched, which we have
found to be the case based on our DFT calculations.

The coercive field for this puckering transformation is
0.18$\times10^{7}$~V/cm for SnS and 0.51$\times10^{7}$~V/cm for
GeSe. Moreover, we found that applying an electric field in $\hat{y}$
at 0.29$\times10^{7}$~V/cm (0.80$\times10^{7}$~V/cm) could also
convert the $\hat{d}_{\rm puck}$ from $\hat{x}$ to $\hat{y}$ for SnS
(GeSe).  The coercive field calculated by this method corresponds to
the electric field at which the unfavorable phase becomes unstable and
can be seen as an upper bound for the coercive field of a real
multi-domain material.  This is usually smaller provided that the
domain walls are mobile at that temperature and, according to a recent
work~\cite{qian2016}, the domain wall energy is small for this class
of materials.  Thus, the electrical fields necessary for ferroelectric
switching are clearly achievable in current 2D
experiments~\cite{radisavljevic-NatNanotechnology-6-147-2011}.  The
structures of SnS monolayer under electric fields at which the
puckering orientation switches are plotted in Fig.~\ref{fig:efield}.

Since the two materials possess a spontaneous, reversible polarization
and bistability, they classify as ferroelectrics.  The
configuration-coordinate diagram of GeSe is typical of a ferroelectric
with second-order phase transition at $T=0$ (consistent with the
change in symmetry). The energy curve for SnS has a minimum rather
than a saddle point at $Cmcm$-ML, and therefore resembles a
ferroelectric with first order phase transition, with the peculiarity
that the $Cmcm$-ML structure is stable for all $T>0$.  Recently, based
on Car-Parrinello molecular dynamics simulations, Mehboudi et al.
showed that monolayer monochalcogenides undergo an order-disorder
phase transition~\cite{mehboudi-NL-5b04613-2016}. Hence, since SnS and
GeSe have four degenerate $Pnma$-ML phases, we expect that the average
total polarization goes to zero as temperature approaches $T_{\rm m}$.

\subsection{Band structure}
\begin{figure}
  \centering
  \includegraphics[width=8.6cm]{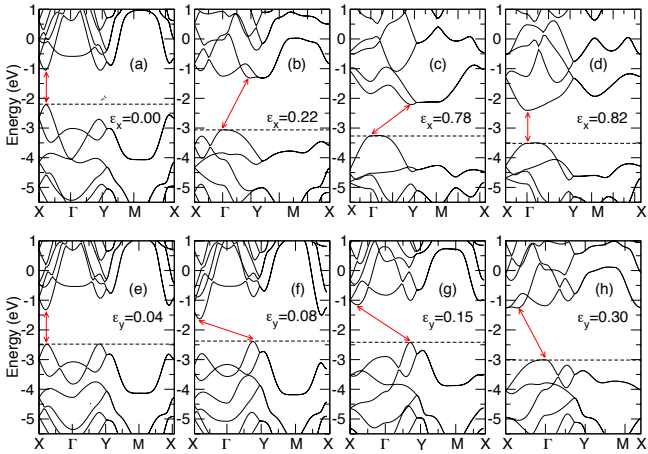}
  \caption{Evolution of GeSe band structure with strains in armchair
    (a) to (d) and zigzag direction (e) to (h). The dotted lines
    locate the valence band maxima. The structure inversion is found
    at $\epsilon_{y}=0.15$. It is apparent that the band structure (b)
    $\epsilon_x=0.22$ (or an uniaxial stress of
    $\sigma_{xx}\sim1.4$GPa) is equivalent to the band structure (h)
    $\epsilon_y=0.30$ (or an uniaxial stress of
    $\sigma_{yy}\sim1.4$GPa) if the $\hat{x}$ and $\hat{y}$ are
    inverted (rotation around $\Gamma$ axis on figures).}
  \label{fig:band2}
\end{figure}
The phase transitions are accompanied by changes of the band structure
and can, therefore, be detected optically.  Representative SnS and
GeSe band structures under uniaxial stress are shown in
Fig.~\ref{fig:band1} and Fig.~\ref{fig:band2}, respectively.  We note
that even though the band gap is underestimated due to our usage of
DFT as the calculation method~\cite{gomesPRB2015}, the dispersion of
the bands is accurately reproduced. Unstrained SnS is an indirect-gap
semiconductor with its valence band maximum located near the X-point
(along the $\Gamma$-X line) and the conduction band minimum near the
Y-point (along the $\Gamma$-Y line).  There are, therefore, two
two-fold degenerate valleys, designated $V_x$ and $V_y$, respectively.
At large strains along $\hat{x}$, SnS transforms to a hexagonal phase
at $\epsilon_x=0.72$ resembling blue phosphorene (Fig.~\ref{fig:band1}
(c))~\cite{mehboudiPNAS2015} and becomes a direct gap at
$\epsilon_x=0.75$. For uniaxial stress along $\hat{y}$ there is a
transition from indirect gap to direct gap at $\epsilon_y=0.02$ (see
Fig.~\ref{fig:band1}(e)), after which the system again becomes
indirect gap.

The band structure of GeSe under uniaxial stress is shown in
Fig.~\ref{fig:band2} (a) to (d) for the $\hat{x}$ and (e) to (h) for
the $\hat{y}$.  Even though unstrained GeSe is a direct-gap
semiconductor, there are also two nearly degenerate conduction band
minima at the $V_x$ and $V_y$ points.  The swapping between the
$\hat{x}$ and $\hat{y}$ of the $Pnma$-ML structure under tensile
stress along the $\hat{y}$ direction occurs at $\epsilon_y=0.15$ and
is in this case accompanied by a loss of the direct bandgap, which
becomes indirect as the structure reverts back into $Pnma$-ML. As
shown in Fig.~\ref{fig:band2}, the band structure (b)
$\epsilon_x=0.22$ is equivalent to the band structure (h)
$\epsilon_y=0.30$ if the $\hat{x}$ and $\hat{y}$ are inverted
(rotation around $\Gamma$ axis on figures). The transition to a
hexagonal phase under tensile stress along $\hat{x}$
($\epsilon_x=0.78$) is also accompanied by an indirect- to direct-gap
semiconductor transition.

\begin{figure*}
  \centering
  \includegraphics[width=17cm]{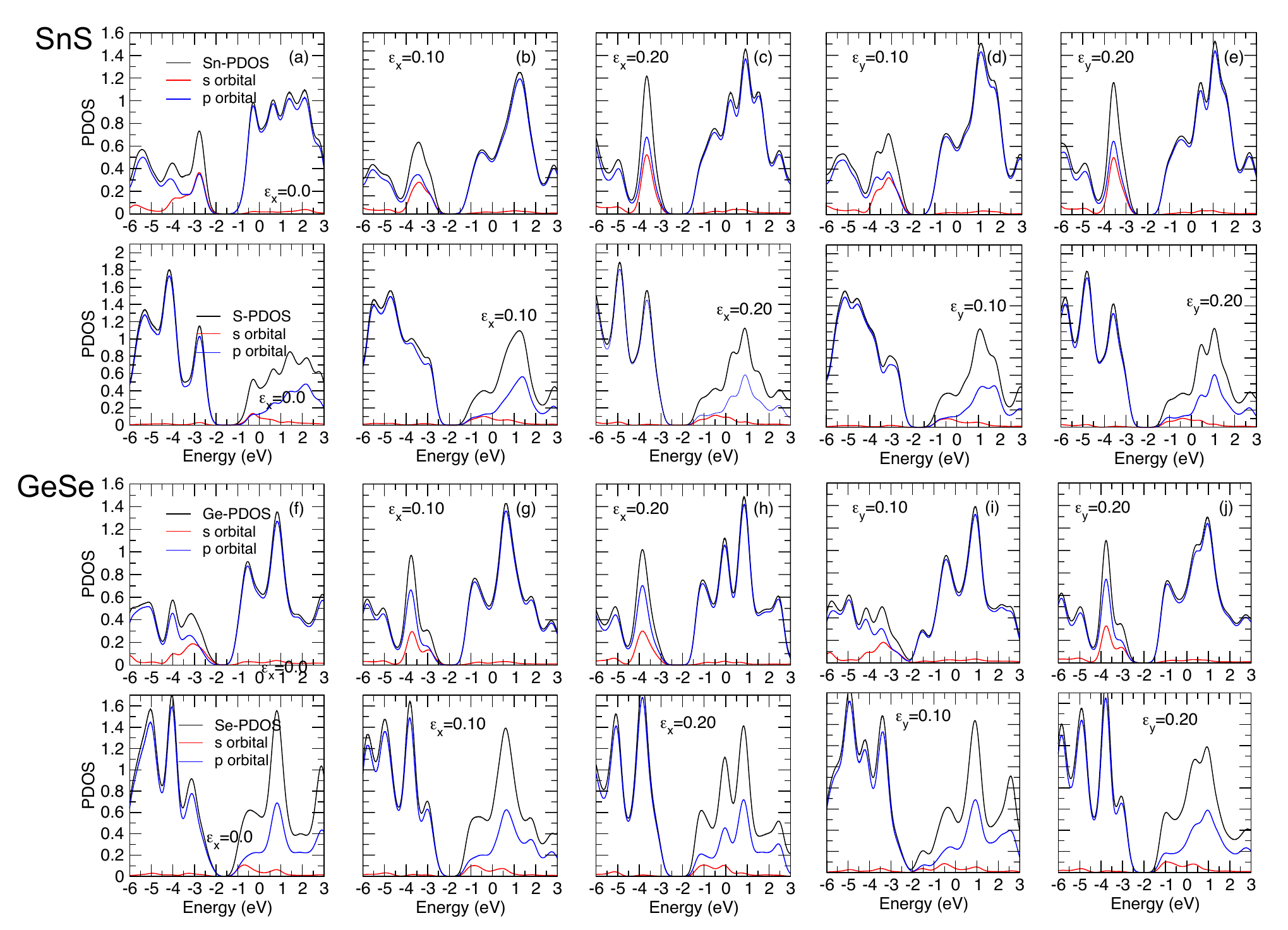}
  \caption{Projected density of states (PDOS) of SnS (a) to (e) and GeSe (f) to (j) for different
    strains. The top panels are PDOS of Sn (Ge) atom and the bottom panels
    are PDOS of S (Se) atom.}
  \label{fig:PDOS-all}
\end{figure*}

In addition, we calculated the projected density of states for SnS and
GeSe for various strains (Fig.~\ref{fig:PDOS-all}).  The trends of the
evolution of PDOS of GeSe and SnS with increasing strain are similar.
Specifically, the relative contributions of the $p$-orbitals for Sn
and Ge atoms at energies close to the maximum valence band increases
with increasing strain.

\begin{figure}
  \centering
  \includegraphics[width=8.6cm]{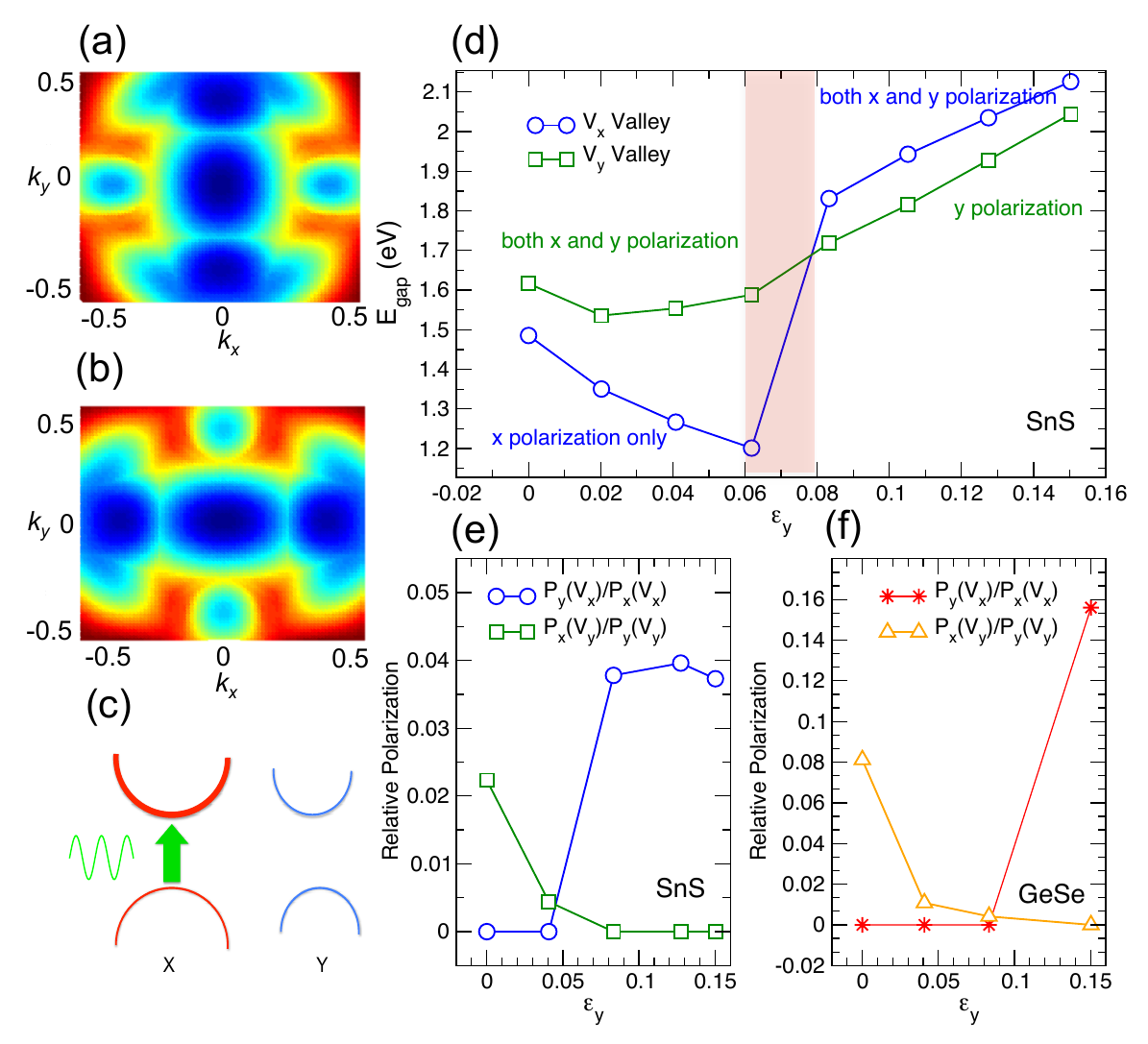}
  \caption{Band gap surfaces (a) $\epsilon_x=0.22$ and (b)
    $\epsilon_y=0.27$ demonstrate the valley swapping. (c) Schematic
    selective valley polarization.  (d) Evolution of the bandgap and
    (e, f) relative polarization under uniaxial stress along
    $\hat{y}$, highlighting the phase transition.  Under small strain,
    the direct transition at $V_x$ is only visible under incident
    $x$-polarized light, while the $V_y$ transition is visible under
    both incident $y$ and $x$ (with a small coupling) polarized
    light.} 
  \label{fig:bandgap}
\end{figure}

The selection of valleys $V_x$ or $V_y$ can be achieved by at least
two different optical methods: (i) using the fact that the direct gap
is different at the two valley pairs; or (ii) using the optical
selection rules. The direct transitions at the $V_x$ and $V_y$ valleys
have different energies, provided there is a means to identify the
orientation of the crystal (Fig.~\ref{fig:bandgap}).  We plot the
energy difference between valence and conduction band of SnS as
functions of in plane wave vectors shown in Fig.~\ref{fig:bandgap} (a)
and (b). It can be seen that the gap surface of $\epsilon_x=0.22$
(Fig.~\ref{fig:bandgap} (a)) is equivalent to $\epsilon_y=0.27$
(Fig.~\ref{fig:bandgap} (b)) but rotated 90$^\circ$. It is evident
that under uniaxial stress in $\hat{y}$ the bands have rotated in the
Brillouin Zone, i.e. the $V_y$ valley effectively becomes the $V_x$
valley after passing the transition point of $\epsilon_{y}$=0.08.

\subsection{Transition probabilities}
Using linearly polarized light to select the valleys $V_x$ or $V_y$
provides an additional method to detect the phase transition
optically. The interband transition probability at a given wave vector
$\bf k$ is given by~\cite{rodin-PRB-93-045431, xiao-PRL-108-196802-2012}
\begin{equation}
P_i({\bf k})\propto\left|\frac{m}{\hbar}\left\langle c({\bf k}) \left|\frac{\partial H}{\partial k_i}\right|v({\bf k})\right\rangle\right|^2,
\end {equation}
where $i$ is the direction of the light polarization, $c(\bf k)$ is
the conduction band wave function, $v(\bf k)$ is the valence band wave
function, and $H$ is the Hamiltonian.  Alternatively, one can relate
the transition probability to the dipole moment between the initial
and the final bands:
$\langle \mathbf{c}|\hat p_{x/y}|\mathbf{v}\rangle$, where the
momentum direction corresponds to the light polarization. For the
transition to be allowed, the dipole moment must not vanish. It is
possible to determine whether it is finite or not using the symmetry
of the bands and the momentum. Since the dipole moment is computed by
integrating the product of the initial and final wave functions, and
the momentum, it is nonzero only if this product
($\propto \mathbf{c}^\dagger (\mathbf
{r})\partial_{x/y}\mathbf{v}(\mathbf{r})$)
is not odd with respect to any of the axes. In other words, the
integrand must remain unchanged under every symmetry transformation of
the space group characterizing the crystal.

We used our \emph{ab initio} results to calculate the transition
probabilities.  For unstrained SnS, $\hat{y}$-polarized light
populates only the $V_y$ valleys, as there is no coupling between the
valence and conduction band at $V_x$ in the $\hat{y}$ direction (see
Fig.~\ref{fig:bandgap} (e)). As shown by
Ref. ~\cite{rodin-PRB-93-045431}, the conduction band, valence band,
and the $p_x$ have a same irreducible representation. Consequently,
the direct product of these quantities results in a non-vanishing
transition probability coupling. On the other hand,
$\hat{y}$-polarized light cannot excite $V_x$, as it possesses
different representation.  $\hat{x}$-polarized light can populate both
$V_x$ and $V_y$ but it populates predominantly the $V_x$ valleys, with
$P_x(V_x)/P_x(V_y)\sim40$.  Similar behavior is observed in GeSe with
a smaller selective valley polarization ratio. For instance, with
linearly $\hat{x}$-polarized light the selective valley polarization
ratio was found to be $P_x(V_x)/P_x(V_y)\sim15$.  The schematic valley
polarization is shown in Fig.~\ref{fig:bandgap}~(c).

The evolution of local gap $V_x$ and $V_y$ of SnS under stress in the
$\hat{y}$ direction is shown in Fig.~\ref{fig:bandgap}(d). We see that
there is an abrupt change in $V_y$ gap near the transition point
$\epsilon_y=0.08$. We also plot the relative polarization
$P_y(V_x)/P_x(V_x)$ and $P_x(V_y)/P_y(V_y)$ as a function of axial
strain $\epsilon_y$, shown in Fig.~\ref{fig:bandgap}(e) and (f) for
SnS and GeSe, respectively. As we discussed earlier,
$\hat{x}$-polarized light populates predominantly the $V_x$ valleys
but there is still a small transition probability at $V_y$ when
$\hat{x}$-polarized light is used. The absorption threshold for
$\hat{x}$-polarized light has an abrupt change near $\epsilon_y=0.08$
($\epsilon_y=0.15$ for GeSe), when the phase transition takes
place. However, the absorption edge for $\hat{y}$-polarized light
changes smoothly.

Before the transition point, the structure has a mirror symmetry
inverting $\hat{y}$, and the $V_y$ valleys can be populated using
polarized light along $\hat{y}$ and $\hat{x}$ (the latter with a very
small coupling).  However, after the transition point, the puckering
direction is rotated to be in the $\hat{y}$, and the reflection
symmetry in $\hat{y}$ is broken, whereas a reflection symmetry emerges
in $\hat{x}$. As a result, $V_x$ can be excited by both $\hat{x}$ and
$\hat{y}$ polarized light after the transition takes place.  We have
therefore demonstrated how optical transitions can be used to detect
the orientation of the structure which determines valley
configurations.

In summary, we have used first-principles calculations to demonstrate the
potential of group-IV monochalcogenide monolayers as functional
materials for information storage.  This strategy, demonstrated using
SnS and GeSe as prototypes, relies on the metastability and the
possibility of switching the polarization direction using stress or
electric field, creating a binary memory device.  Comparing these
prototype materials, SnS differs from GeSe because it has a stable
centrosymmetric phase which, at $T=0$, is close in energy to the
$Pnma$-ML phase.

Due to their peculiar band structures, both SnS and GeSe could in
principle be used as functional materials for memory devices that can
easily be interfaced with valleytronics logic.  Valleytronics is based
on the concept that the valley index can potentially be used to store
information for subsequent logic operations, equivalent to spin in
spintronics.  However, in most valleytronics materials the information
can be considered non-volatile only up to the timescale defined by
inter-valley scattering processes, which are ubiquitous in real
materials.  Structural changes, used to store information in phase
change memory devices, take place on a timescale orders of magnitude
longer.  Materials such as SnS and GeSe can be used to convert
information stored as structural phase into information stored as
valley index.  One possibility is for example by using near-bandgap
light that excites only the pair of valleys corresponding to the
lowest energy exciton.  The subsequent electronic state will have
electron-hole pairs with momentum $(\pm k_x,0)$ or $(0,\pm k_y)$,
depending on the structure orientation.  This valley state can be
transmitted onto a valley-filter~\cite{gunlycke-PRL-106-136806}.
Alternatively, if coupled to a polarized light detector, the
polarization switching can be detected optically taking advantage of
valley-dependent direction of the linear polarization of the
luminescence~\cite{rodin-PRB-93-045431}.

\begin{acknowledgments}
  P.Z.H. is grateful for the support of the Physics and Mechanical
  Engineering Departments at Boston University, the hospitality of the
  NUS Centre for Advanced 2D Materials and Graphene Research Centre
  where this work was initiated, the support of the Materials Science
  and Engineering Innovation Grant and the Boston University High
  Performance Shared Computing Cluster. A.C. acknowledges support by
  the National Research Foundation, Prime Minister Office, Singapore,
  under its Medium Sized Centre Programme and CRP award ``Novel 2D
  materials with tailored properties: beyond graphene'' (Grant No.
  R-144-000-295-281). D.K.C. is grateful for the hospitality of the
  Aspen Center for Physics which is supported by NSF Grant
  $\#$PHY-1066293, and of the International Institute for Physics of
  the Federal University of Rio Grande do Norte, in Natal, Brazil,
  where some of this work was completed.  HSP acknowledges the support
  of the Mechanical Engineering Department at Boston University. We
  thank Alex Rodin for helpful comments and discussions.
\end{acknowledgments}

\bibliography{biball}

\end{document}